\documentclass[final,5p,times,twocolumn]{elsarticle}

\usepackage{booktabs}
\usepackage{tabularx}
\usepackage{float}
\usepackage{color,soul}
\usepackage{amssymb}
\usepackage{gensymb}

\usepackage{amsmath,array,graphicx}

\usepackage{pdfcolparallel}

\makeatletter
\def\@author#1{\g@addto@macro\elsauthors{\normalsize%
    \def\baselinestretch{1}%
    \upshape\authorsep#1\unskip\textsuperscript{%
      \ifx\@fnmark\@empty\else\unskip\sep\@fnmark\let\sep=,\fi
      \ifx\@corref\@empty\else\unskip\sep\@corref\let\sep=,\fi
      }%
    \def\authorsep{\unskip,\space}%
    \global\let\@fnmark\@empty
    \global\let\@corref\@empty  
    \global\let\sep\@empty}%
    \@eadauthor={#1}
}
\makeatother

\usepackage[table]{xcolor}
\definecolor{mycolor}{RGB}{255,255,204}

\usepackage{lineno}				

\usepackage{hyperref}
\hypersetup{
    colorlinks=true,
    linkcolor=blue,
    filecolor=magenta,      
    urlcolor=cyan,
}
\modulolinenumbers[5]			

\usepackage{subcaption,graphicx}
\usepackage{makecell}

\graphicspath{{images/}}

\usepackage[labelsep=endash]{caption}

\journal{Astronomy and Computing}

\biboptions{authoryear}		

\newcommand*{\pycloudy}{\textsc{PyCloudy}}
\newcommand*{\altpycloudy}{{\mdseries\pycloudy}}

\newcommand*{\pycross}{\textsc{PyCross}}
\newcommand*{\altpycross}{{\mdseries\pycross}}

\newcommand*{\cloudy}{\textsc{Cloudy}}
\newcommand*{\altcloudy}{{\mdseries\cloudy}}

\newcommand*{\shape}{\textsc{Shape}}
\newcommand*{\altshape}{{\mdseries\shape}}

\begin{document}
\begin{frontmatter}
\title{Introducing \textsc{\altpycross{}}: \textsc{\altpycloudy{}} Rendering Of \textsc{\altshape{}} Software \\for pseudo 3D ionisation modelling of nebulae}

\author[add1,add2]{K. Fitzgerald\fnref{fn1}\corref{cor1}}
\cortext[cor1]{Corresponding author}
\fntext[fn1]{\href{mailto:kfitzgerald@ait.ie}{kfitzgerald@ait.ie} (K. Fitzgerald)} 
	
\author[add3]{E.J. Harvey\fnref{fn2}}
\fntext[fn2]{ \href{mailto:E.J.Harvey@ljmu.ac.uk}{E.J.Harvey@ljmu.ac.uk} (E.J. Harvey)} 

\author[add1]{N. Keaveney\fnref{fn4}}
\fntext[fn2]{\href{mailto:N.KEAVENEY2@nuigalway.ie}{N.KEAVENEY2@nuigalway.ie} (N. Keaveney)} 

\author[add1]{M.P. Redman\fnref{fn3}}
\fntext[fn2]{\href{mailto:matt.redman@nuigalway.ie}{matt.redman@nuigalway.ie} (M.P. Redman)} 

\address[add1]{Centre for Astronomy, School of Physics, National University of Ireland - Galway, University Road, Galway, H91 CF50, Ireland}
\address[add2]{Permanent address: Athlone Institute of Technology, Dublin Road, Bunnavally, Athlone, Co. Westmeath, N37 HD68, Ireland}
\address[add3]{Astrophysics Research Institute, Liverpool John Moores University, Liverpool, L3 5RF, UK}

%


\begin{abstract}
Research into the processes of photoionised nebulae plays a significant part in our understanding of stellar evolution. It is extremely difficult to visually represent or model ionised nebula, requiring astronomers to employ sophisticated modelling code to derive temperature, density and chemical composition. Existing codes are available that often require steep learning curves and produce models derived from mathematical functions.  In this article we will introduce \altpycross{}: \altpycloudy{} Rendering Of {\sc Shape} Software. This is a pseudo 3D modelling application that generates photoionisation models of optically thin nebulae, created using the \altshape{} software. Currently \altpycross{} has been used for novae and planetary nebulae, and it can be extended to Active Galactic Nuclei or any other type of photoionised axisymmetric nebulae. Functionality, an operational overview, and a scientific pipeline will be described with scenarios where \altpycross{} has been adopted for novae (V5668 Sagittarii (2015) \& V4362 Sagittarii (1994)) and a planetary nebula (LoTr1).  Unlike the aforementioned photoionised codes this application neither requires any coding experience, nor the need to derive complex mathematical models, instead utilising the select features from \altcloudy{}\,/\,\altpycloudy{} and \altshape{}.   The software was developed using a formal software development lifecycle,  written in Python and will work without the need to install any development environments or additional python packages.  This application, \altshape{} models and \altpycross{} archive examples are freely available to students, academics and research community on GitHub for download. 
\end{abstract}

\end{frontmatter}

\section{Introduction}
Some of the most iconic images of modern astronomy are of nebulae, made visible through emission lines in photoionized gas. The study of such photoionised gas has played a major role in understanding a number of physical processes ranging from atomic physics to stellar evolution theories. Visible light from nebulae originates directly from photoionised gas at an equilibrium temperature of approximately $10^3$--$10^4$\,K \citep{2006AstroBook}.  As PN central stars have very high temperatures, the gas is excited by ultraviolet radiation from a central star with a very high surface temperature.  For nova and supernova remnants, shockwaves in the initial explosion flash ionises the gas.  In both cases, recombinations into any state other than the ground state (which produce UV photons that are promptly reabsorbed) lead to escape of photons from the nebula rendering them visible particularly at optical wavelengths. Nebulae whose exterior boundary occurs at the outer edge of the gas are referred to as ``\textit{matter-bound}".  In contrast, ``\textit{radiation-bound}" nebulae occur when the hydrogen ionisation front refines the outer boundary, as seen in visible light, since ionisation fronts can also be imaged in the UV and IR.  In the latter case, more gas will be present beyond the visible boundary, and may emit in longer wavelengths in a `photon dominated region'. Finally, reflection nebulae are caused by scattered starlight (e.g. around the Pleiades star cluster) rather than photoionisation. In this paper, the focus is restricted to optical emission from inside photoionised nebulae.  While the focus is on optical emission, the software can of course be used for UV--IR emission.

Planetary nebulae (PNe) spectra consist predominately of emission lines, a result of ionisation from UV radiation emitted from a post asymptotic giant branch (AGB) star and interaction of electrons within the nebula.  These lines are categorised as either recombination lines: formed when an ion and an electron combine, leading to a cascade of the electron down to the ground state, or collisionally excited lines: formed when  electrons collide with atoms or ions within the gas, and excite them. When these excited atoms or ions revert to a lower level, they will emit a photon.  Lines known as \textit{`forbidden lines'} do not occur under terrestrial conditions, and are denoted by square brackets, (e.g [N\,{\sc ii}], [O\,{\sc iii}] etc.).  Abundances of elements within a nebula are generally expressed as a ratio in comparison to the intensity hydrogen beta line (i.e. H$\beta$ 4861$\AA$). Spectroscopic analysis can determine chemical composition, density, velocities and temperatures of regions within a nebula.

An important and current problem in abundance determinations in PNe and novae exist as a discrepancy between the abundances determined from forbidden lines and those determined from analysis of recombination lines.  The ratio of these deviations is defined by the abundance discrepancy factor (ADF).  Gaseous nebula have an ADF in the range of 2--3, in PNe this can be as high as 70 \citep{2015Torres-Peimbert}. It has been argued that the ADF in PNe is  due to: (i) chemical inhomogeneities (fluctuations or variations), (ii) temperature inhomogeneities in chemically homogeneous objects and (iii)  the destruction of solid bodies inside PNe that produce cool and high-metallicity pockets \citep[see][and references therein]{2017Peimbert_etal}.  \cite{2015Corradi_etal} investigated the ADF problem in three PNe (Abell 46, Abell 63 and Ou5) with close binary central stars (orbital periods of $\sim10$\,hrs) and may have passed through common envelope evolution and showed all to have a  large ADF.  \citeauthor{2015Corradi_etal} do not solely credit the high ADF to close binaries and debate that the origins of the metal rich gas present in the nebula may also be influenced by nova-like outbursts, planetary material and the destruction and engulfment of circumbinary Jovian planets \citep{2017BoyleRedman, 2018PhDT_Boyle}.
Photoionisation modelling of ejected nova shells during their nebular stage of evolution can contribute to estimates of the total mass and abundances of heavy elements ejected during nova events \citep{2011Atypical}.

Research into the processes of photoionised gas plays a huge part in our understanding of stellar evolution, which is why so much effort has been put into creating tools to help visualise it.  Through the analysis of such lines, accompanied by photoionisation models, the physical condition and chemical abundances of PNe and novae can be further understood.  In order to create a visual representation of an ionised nebula, a modelling code to derive temperature, density, chemical composition, and other physical quantities is required. To aid in defining the initial values for 3D photoionisation models interactive databases, such as PyNeb \citep{2015PyNeb}, are available that aid the user in deriving physical conditions of a nebula based on observed spectral line ratios.
 
Modelling non-stellar objects beyond our solar system is difficult as 3D volumetric objects are observed as 2D projections.  Currently there are a number of codes available that use various analytical or statistical techniques for the transfer of continuum radiation, mainly under the assumption of spherical symmetry.  Increased computing power has further enhanced the development of  photoionisation codes, allowing for the construction of more complex models reconstructed into 3D.  However in some cases, the assumption of spherical symmetry has been retained \citep{2005Photoionisation}.  Two leading contributions to this area include MOCASSIN and \altcloudy{}.

MOCASSIN (MOnte CArlo SimulationS of Ionised Nebulae), described in \cite{2003MOCASSIN},  is a code designed to build realistic models of photoionised nebulae having arbitrary geometry and density distributions, with both the stellar and diffuse radiation fields treated self-consistently.  This Monte Carlo approach was developed to provide a fully 3D modelling tool capable of dealing with asymmetric and/or inhomogeneous nebulae, as well as, if required, multiple, non-centrally located exciting stars.  The time taken to run/converge simulations and renderings on a standard desktop ranges form a number of hours to days.  An alternative to this is the pseudo 3D technique of \altcloudy{}\footnote{Cloudy is available under general use under an open source licence: \url{http://www.nublado.org}} \citep{1998CLOUDY90, 2000CLOUDY00, 2013Cloudy}.  \altcloudy{} is a large-scale spectral synthesis code designed to simulate physical conditions within an astronomical plasma and then predict the emitted spectrum.  \altcloudy{} 3D\footnote{Cloudy3D is available online \url{https://sites.google.com/site/cloudy3d/}} is an IDL library to compute pseudo 3D photoionisation models by interpolating radial profiles between several 1D \altcloudy{} models. Users can generate emission line ratio maps, position-velocity (PV) diagrams, and channel maps, once an expansion velocity field is given.  It is significantly faster than full 3D photoionisation codes, allowing users to explore a wide space of parameters quickly.  \cite{2013pyCloudy} 
developed \altpycloudy{},\footnote{Python wrapper for cloudy.  Available online: \url{https://sites.google.com/site/pycloudy/}} a Python library that handles input and output files of the \altcloudy{} photoionisation code.  \altpycloudy{} can also generate pseudo 3D renderings from various runs of the 1D \altcloudy{} code.

Typically only 1D velocity information is available from the Doppler shifting of lines along the lines of sight.  In special cases, if velocities are high enough and the sources are close enough, then {\it proper motions} can be observed, where, over time, the nebula features can be observed to advance on the plane of the sky. Regarding novae and nebulae, photographic images provide a 2D integration of the emission and absorption along the line-of-sight, but the depth information is flattened.  Additional information may be assumed based on the symmetry and orientation of the object being observed.  Depth information may also be extrapolated from velocity fields, e.g. mapping between velocity and position in radially expanding nova shells. This requires prior knowledge of the properties of the object being modelled.  Alternative methods of modelling are required when theoretical or observational constraints are insufficient, one such example of this is \altshape{}\footnote{\textsc{\altshape{}} is available online: \url{http://www.astrosen.unam.mx/shape/index.html}}\citep{2011SHAPE}.

\begin{figure}[!h]
	\centering
	{\includegraphics[width=3.5in]{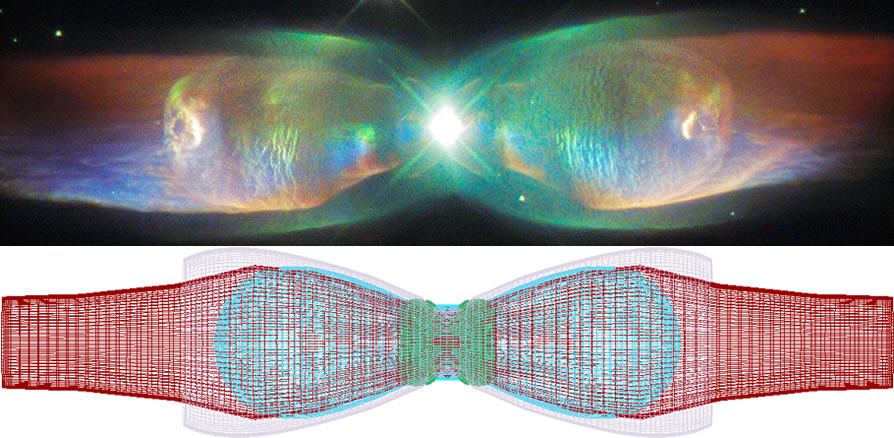}} %
	\caption[PNe M2-9]{PNe M2-9:  Top image captured from NASA Hubble telescope and processed by Judy Schmidt\footnote{},  filters used: [S\,\textsc{ii}], [O\,\textsc{iii}] and H\,$\alpha$.  Bottom image \textsc{Shape} 3D model.  \textit{Top image reproduced from \url{https://www.nasa.gov/feature/goddard/hubble-sees-the-wings-of-a-butterfly-the-twin-jet-nebula} }.  Refer to \cite{M2-9_Doyle_2000} for more information regarding the morphology of M2-9.}
	\label{fig:m29}
\end{figure}

\altshape{} is a morpho-kinematic modelling and reconstruction tool for astrophysical objects.  Users bring any knowledge of the structure and physical characteristics of the source (e.g. symmetries, overall appearance, brightness variations) to construct an initial model which can be visualised. The model can be compared to observational data allowing for interactive and iterative refinement of the model.  Once all necessary physical information are reflected in the model, its parameters can be automatically optimised, minimising the difference between the model and the observational data. The final model can then be used to generate various types of graphical output, \autoref{fig:m29} shows a \altshape{} model of PN M2-9 under an image of M2-9 acquired by NASA (refer to \autoref{fig:m29}).  Recent examples where \textsc{\altshape{}} has been employed to model PNe and novae include: \cite{2014Clyne} create a kinematical model of PN MyCn 18, utilising expansion velocities of its nebular components by means of position velocity (PV) arrays, to ascertained the kinematical age of the nebula and its components. \cite{2016Harvey_etal} modelled the Firework nebula and discovered that the shell was cylindrical and not spherical as previously believed. The lower velocity polar structure in this model gave the best fit to the spectroscopy and imaging available. \cite{2019Sophia} presented a  morpho-kinematical model of PN HB4 using new Echelle spectroscopic data and high-resolution HST images.  \citeauthor{2019Sophia} concluded that HB4 had an absolute mean expansion velocity of $14\,\rm{km\,s}^{-1}$ along the line of sight and proposed that the central part of the nebula consists of a binary system that has a Wolf--Rayet (WR) type companion evolved through the common-envelope channel. Refer to the following for more examples of \altshape{} software used to model PNe and nova: \cite{2016Akras_etal, 2017Starfish, 2017NGC6302, 2018PNe_ExtendedStructure}.

\footnotetext{\url{https://photographingspace.com/apod-judy-schmidt/}}

High resolution imaging may detect shapes, however better understanding of the three dimensional structure is achieved when this data is coupled with spatially resolved, high resolution spectra to determine the kinematics of the gas within the nebula.  To this end we present a new application and pipeline that uses \altshape{} software to create 3D models with density, velocity and temperature properties.  The output of this is a data-cube which is processed by \altpycross{} to generate \altcloudy{}  photoionisation models of nebulae.  This is achieved using an intuitive interactive graphical user interface (GUI) that does not require programming experience, execution scripts or the need to install additional compilers or libraries.  The paper outline is as follows: Section \ref{Section2PyCross} discusses the development, installation, functionality and operational overview of \altpycross{}.  Section \ref{Section3Novae} demonstrates \altpycross{}' scientific pipeline outlining its use on novae V5668 Sagittarii (2015) and V4362 Sagittarii (PTB 42).  Section \ref{Section4PNe} again demonstrates the pipeline for PN LoTr1.
Section \ref{Section5Conslusion} consists of our discussion and conclusions.

\begin{figure*}[!t]
	\centering
	{\includegraphics[width=\textwidth]{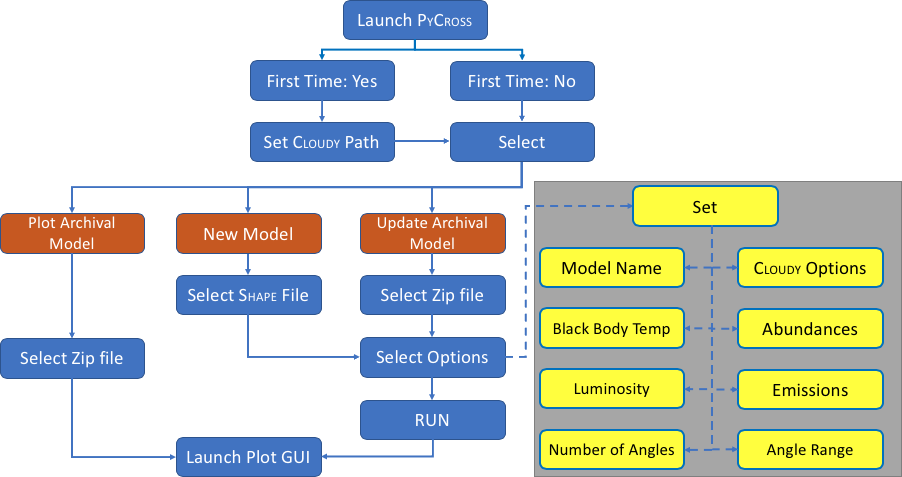}} %
		\caption[PyCross Operational Flowchart GUI]{\textsc{PyCross} Operational Flowchart.  Steps for generating new models, updating archived models by modifying \textsc{Cloudy} parameters and visualising current and archived models.  A high-level overview of the path to generate a new model is illustrated in \autoref{fig:GUI_Overview}.}
	\label{fig:PyCross_FlowChart}
\end{figure*}

\section{\textbf{\textsc{PyCross} }} \label{Section2PyCross}
The steep learning curves encountered when installing, understanding and utilising specialist software are further compounded when astronomers are often required to develop/code software for specific tasks. \cite{2015_Astro_Software_Survey} carried out a survey, between December 2014 and February 2015, that focused on the use of software and the software skills of  participants in the astronomy community worldwide.  Participants consisted of 1142 astronomers ranging from students to postdocs and research scientists.   All survey participants responded ``yes" when asked if they used software as part of their research; 11\% of responders said they used software developed by others; 57\% used software developed by themselves and others; while 33\% said they used software they developed themselves for specific purposes, as there was no software readily available.  This research also revealed the open source language, Python, to be the programming language of choice.  

With advances in techniques, hardware and software it is of no surprise the number of astronomy software applications made available has increased considerably over the last decade. More recently \cite{FITZGERALD2019} developed a standalone application (written in Python) for plotting photometric Colour Magnitude Diagrams (CMDs) using object orientated programming (OOP), a formal Software Design Lifecycle (SDLC) and Test Driven Development (TDD). This stand alone application worked ``\textit{out of the box}" required no installation of any additional software to function and emphasised the importance of quality and standards when developing software for astronomy.  A SDLC  defines a structured sequence of stages in software engineering to develop the intended software product, a TDD approach relies on a shorter development cycle, where requirements become specific test cases that the software must pass and OOP allows developers to   break software projects down into smaller, more manageable modular problems, one object at a time.

\subsection{Software development}
\altpycross{} was developed and tested on OSX using the PyCharm IDE (integrated development environment) 2018 Community Edition, Python 2.7 with later iterations using Python 3.7.  The free community edition of PyCharm offers usage of both testing frameworks\footnote{PyCharm Testing Frameworks: \url{https://www.jetbrains.com/help/pycharm/testing-frameworks.html}} and code analysis tools\footnote{PyCharm Editions Comparison: \url{https://www.jetbrains.com/pycharm/features/editions_comparison_matrix.html}}.  There were 24 iterations of this application before the code was ``\textit{Frozen}", for this release, allowing for the creation of a single executable file that can be distributed to users\footnote{Currently this application is only available as an executable on MAC OS and Windows.}. A frozen application contains all the code and any additional resources required to run the application, and includes the Python interpreter that it was developed on.  The major advantage for distributing applications in this manner is that it will work immediately, without the need for the user to have the required version of Python (or any additional necessary libraries) installed on their system.  A disadvantage of generating a single file is that it will be large (approximately 183\,MB), as all necessary libraries are incorporated, the increase in file size is acceptable when considering other issues, for example ease of installation, running, and portability to other platforms.
The Agile SDLC and TDD approaches employed in \cite{FITZGERALD2019} were adapted in the development of this software, refer to this text for a comprehensive review of employing specific SDLC and TDD approaches in astronomy.  The remainder of this section will outline installation, features and operational overview of the functionality when using \textsc{PyCross}

\begin{figure*}[!th]
	\centering
	{\includegraphics[width=\textwidth]{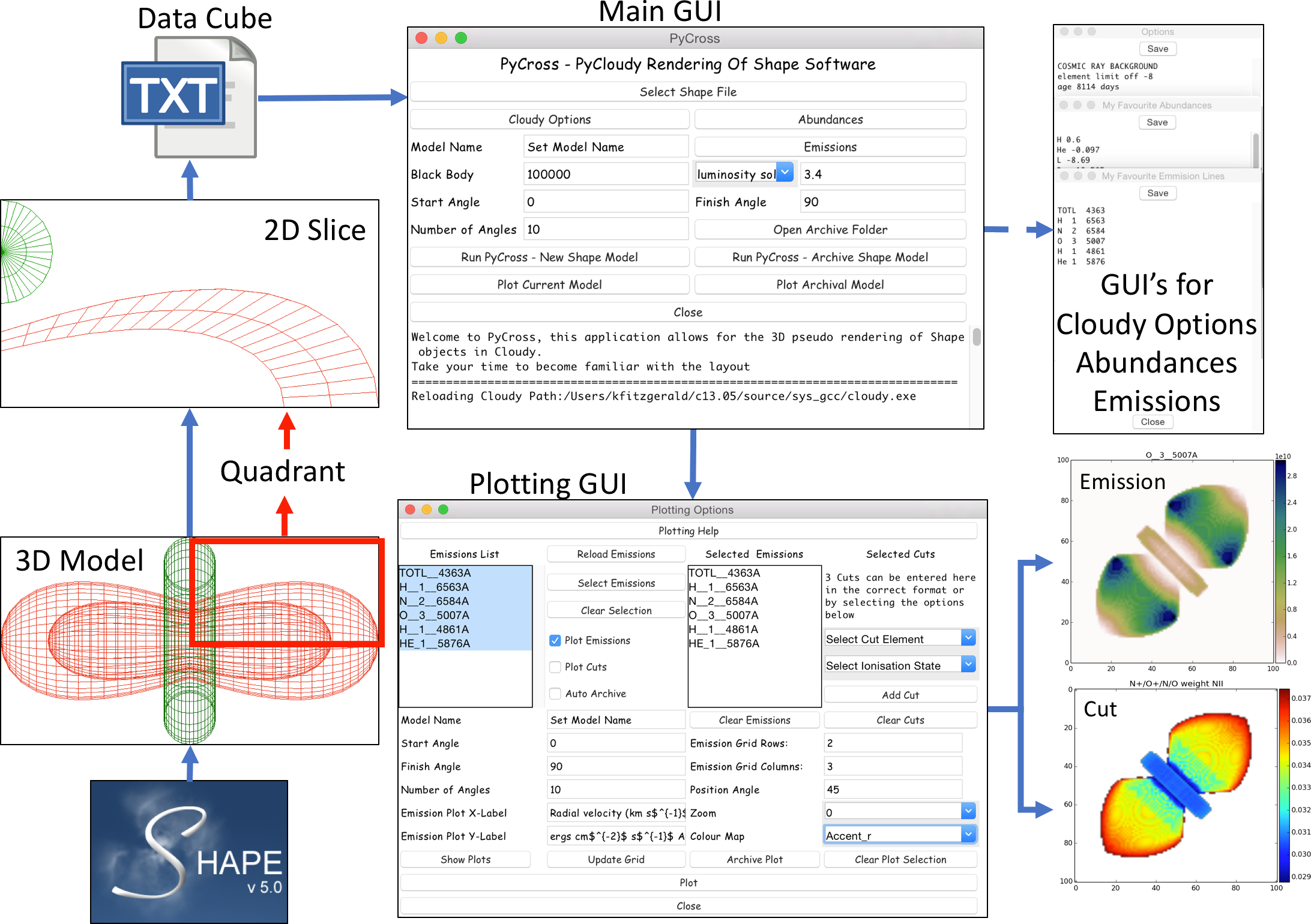}} %
		\caption[High-level operational overview of \altpycross{}]{A high level operational overview following the path for creating a new model outlined in \autoref{fig:PyCross_FlowChart}.  A new 3D \textsc{Shape} model is created with appropriate structure, densities and velocities.  A quadrant of the overall model is then sliced to produce a data cube, in the format of a text file \textbf{(\autoref{shapetextfile})}, which is inputted into \altpycross{}.  \textsc{Cloudy} options/parameters are then specified and the model is run.  \altpycross{}'s console will update informing the user of the progress.  Once completed the user opens the plotting interface to plot the photoionisation and cut models (see bottom right of diagram) at different angles  and inclinations.  Colour map used for emission is gist\_earth\_r, one of 71 available to users.  The model used here was designed as a theoretical nova with cosmic ray background and an approximate age of 8114 days, blackbody effective temperature of $50\times10^3$\,K, luminosity of $3.6\,\mathrm{L_\odot}$ and equatorial expansion velocity of $500\,\rm{km\,s^{-1}}$ and position angle of 45\degree.  This model is available online at \url{https://github.com/karolfitzgerald/PyCross_OSX_App}, filename: \textit{Demo1.zip}.  A scientific approach for determining optimum \textsc{Cloudy} parameters is outlined in \autoref{fig:pipeline}}
	\label{fig:GUI_Overview}
\end{figure*}

\subsection{ \textsc{\altpycross{}} functionality}\label{pycrossfunctionality}
\textsc{PyCross} was comprehensively designed to allow for intuitive usability while also providing automated features allowing the user to be more productive.    A high-level operational overview of \textsc{PyCross} functionality is outlined in the operational flowchart in \autoref{fig:PyCross_FlowChart}. Then,  \autoref{fig:GUI_Overview} follows the path for generating  \textsc{PyCross} photoionisation rendering of a 3D \textsc{Shape} model utilising the application's two main GUIs, for creating and plotting models.  Smaller windows are used to manage any \textsc{Cloudy} preferences.   

\subsubsection{Installation \& directory setup} As previously discussed, this application is ``\textit{frozen}" allowing for the creation of a single executable which contains all code and any additional python libraries to run. The only requirement is that \textsc{Cloudy} is pre-installed on the user's system prior to running this programme.

A new directory structure is created in the root folder of the users computer when \textsc{PyCross} is loaded for the first time.  This folder is named ``\textit{PyCrossArchive}" and is the root/destination for all work generated by the application:  the following folders and files are generated upon startup and modified during the execution of a model:

\begin{itemize}
	\item \textit{\textbf{Model-Name-Timestamp.zip}}: \textsc{PyCross} models are automatically saved using the name assigned followed by a time-stamp of when they are created.  The purpose of this is as follows:  1) To archive all models, this allows users the opportunity to review any change in parameters and resulting models generated. 2) A basic model can generate in excess of 100 files, each approximately 14.4\,MB in size.  While maintained within a well structured directory this can quickly consume disk space when generating a lot of models.  Automatically zipping outputs to a single file  significantly reduces file sizes (to approximately 3\,MB zipped for a 14.4\,MB un-zipped file) while also making it easier to process data at a later stage. For example, when handling these archived models, \textsc{PyCross} automatically extracts the contents of the selected zipped file into a temporary folder, and when finished the contents of the temporary folder are deleted. A time-stamp incorporated into the file name allows users to distinguish between models, if the model name does not change.
	\begin{itemize}
		\item	\textit{\textbf{LogFile.txt}}: A text file that records the parameters used by the user and information relating to the progress of a model as it is being run.  This information is also displayed on the main GUI.  This file can be used at a later stage to compare models based on their parameters but also to recreate models if needed. 
		\item	\textit{\textbf{MakeData}}: This folder contains the output files generated by \textsc{Cloudy} based on the \textsc{Shape} model.
		\item	\textit{\textbf{TempData}}:  This folder contains data generated by \textsc{PyCross} that allows \textsc{Cloudy} to run the selected \textsc{Shape} model.  It is also used to extract archive model data when modifying or plotting archived models.
		 	\end{itemize}

	\item \textit{\textbf{Plots}}: This folder contains the following sub-folders which store the generated plots of a model based on their type.  Each time plots are generated the contents of this folder are deleted and replaced with new plots.
  	\begin{itemize}
		\item  \textit{\textbf{Cuts}}: This folder contains plots of the cuts generated.  Cuts will be plotted based on their element and ionisation state.  An additional N$^+$/O$^+$/N/O weight N\textsc{ii} cut, adapted from \textsc{PyCloudy}, is also plotted and saved here. 
		\item \textit{\textbf{Emissions}}: Folder containing plots of the generated emission simulations.
		\item \textit{\textbf{PlotArchiveInformation.txt}}: This text file is updated automatically each time new plots are generated and contains information relating to the parameters entered by the user in the plotting options GUI, Plot GUI (see \autoref{fig:GUI_Overview}), this feature is useful for tracking differences in plots based on their parameters.
	\end{itemize}
	\item \textit{\textbf{PlotArchive}}: All data generated and stored in the \textit{\textbf{Plots}} folder can be exported here and automatically saved as zip files.  This reduces time recreating plots at a later stage while also keeping track of parameters changed and carried out on Plot GUI:
	\begin{itemize}
		\item \textit{\textbf{Auto Archive}}: If this option is selected in the Plotting Options GUI then all current \& subsequent data generated will automatically be exported to the PlotArchive folder as a zip file.  If the user does not change the model name then the naming convention is updated to include the time generated, this ensures that no work is overwritten or lost. 
		\item \textit{\textbf{Archive Plot}}: There is an archive current data option that can be accessed by clicking this button in the Plotting Options GUI. 
	\end{itemize}
\end{itemize}

\begin{figure}[!htb]
	\centering
	{\includegraphics[width=2.5in]{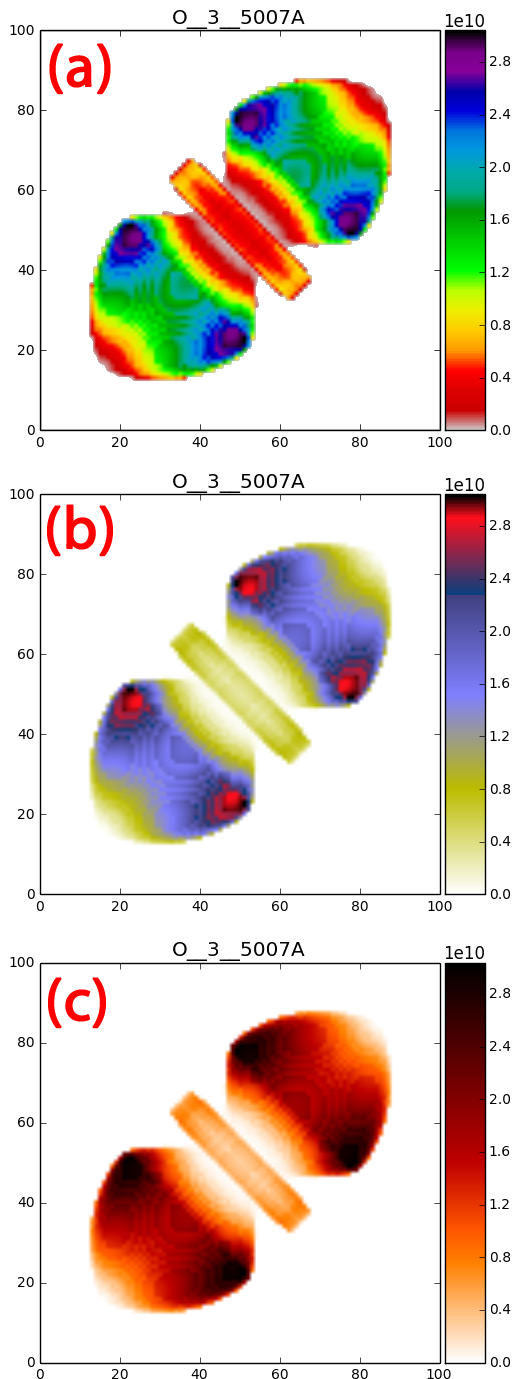}} %
	\caption[Colour Map Options]{Three colour map examples from a choice of 71. The emission model of [O\textsc{iii}]\, 5007$\AA$ is derived from an example model (see \autoref{fig:GUI_Overview}) with a position angle of 45\degree. Colour-bars correspond to the effective temperature of the ionising source, flux units in ergs $\mathrm{s^{-1}}$. Scaled units of physical size are on \textit{x} and \textit{y} axes.}
	\label{fig:colourMapOptions}
\end{figure}

\begin{figure}[!ht]
	\centering
	{\includegraphics[width=\columnwidth]{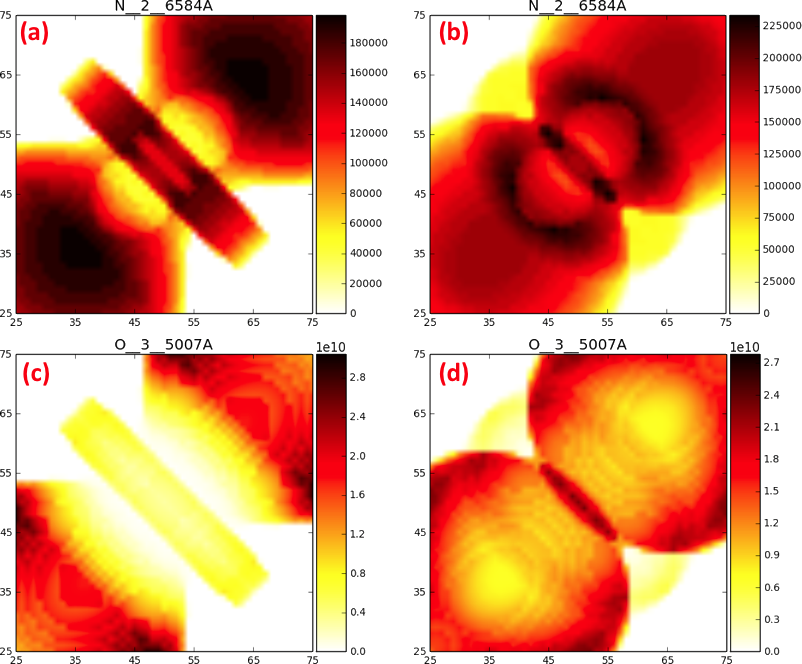}} %
	\caption[Zoom Functionality]{Photoionisation models of [N\,\textsc{ii}]\, 6584$\AA$ (a, b) and [O\,\textsc{iii}]\, 5007$\AA$ (c, d) derived from the theoretical model in \autoref{fig:GUI_Overview}.  Each emission plot initially plotted at an inclination angle of 90\degree (a, c) then at 50\degree (b, d).  Colour-bars correspond to the effective temperature of the ionising source, flux units in ergs $\mathrm{s^{-1}}$.  Plots are zoomed to better visualise features like the torus across different emissions and angles. The colour map used was hot\_r. }
	\label{fig:zoom}
\end{figure}

\subsection{\altshape{} data cube text file }\label{shapetextfile}
	The \altshape{} model is based on observational data and can be informed by a combination of narrow-band imaging, medium-high resolution spectroscopy and polarimetry. If these data types are not all at hand, then they can be used individually with the application of some theory or in certain combinations, given data of good quality. Integral field unit spectroscopy can be a particularly powerful observational tool to aid in the derivation of nebular morpho-kinematics\footnote{To download and for a full description of the \altshape{} software please visit: \url{http://bufadora.astrosen.unam.mx/shape/}}.  In this section the process on how to construct the output data cube from \altshape{} for input into \altpycross{} which is used in the \emph{MakeData} and \emph{TempData} files will be explained.
	
	First users must derived the morphology of the nebular structure under study and assign a fixed or variable density structure.  Next users must create a `cut' of the nebula so it can be represented by a 2D array. To do this the user must place the nebula with its long axis pointing along the x-axis. Then a slice of a quarter of the major-axis of axisymmetric shape is taken, see \autoref{fig:GUI_Overview} (2D Slice), \autoref{fig:pipeline}, panel 2(a) and \autoref{fig:lotr1_Shape} (c). This is done in the \altshape{} 3D module. Users must first go to the object primitive, then take a slice of the nebula by setting $\phi_{min} = 90^{\degree}$, $\phi_{max} = 180^{\degree}$, $\theta_{min} = -91^{\degree}$ and $\theta_{min} = -89^{\degree}$. 
	
	Once users have made an appropriate 2D slice of their nebula they move to the \altshape{} render tab. In this tab the renderer must be set to be \emph{Physical}. The slit and image geometries are set to be square such that they overlap and cover the entirety of the 2D slice, again refer to \autoref{fig:GUI_Overview} (2D Slice), \autoref{fig:pipeline}, panel 2(a) and \autoref{fig:lotr1_Shape} (c) for how the structure should appear in the image render window.  Finally users set up a text file to output the position and velocities along the x, y and z axes (Px, Py, Pz, Vx, Vy Vz) along with the corresponding density and temperature.  This is done under \emph{``Render"} in the Parameters section of \altshape{}.  When users click the Render button a text file will be generated in a folder designated by the user. This text file is then imported directly into \altpycross{} via its GUI. Currently only Px, Py, Pz and density arguments are used as input to the \altpycross{} simulations. Temperature and velocity information will be utilised in the next iteration of \textsc{PyCross}.

\subsubsection{User Interface \& Operational Overview} The Graphical user interface (GUI) is designed to be as intuitive as possible and  consists of two main GUI windows.  The first allows the user to set parameters and run models while the second controls the plotting, these interfaces pass information to each other to enhance efficiency.  Other GUIs allow users to set additional \textsc{Cloudy} options, emission lines and abundances.  The operational flowchart of this application is illustrated in \autoref{fig:PyCross_FlowChart} and a high level overview of the GUIs taking the path of a new model is illustrated in \autoref{fig:GUI_Overview}.  A scientific approach to determine optimum \textsc{Cloudy} parameters is discussed in \autoref{Section3Novae} (see \autoref{fig:pipeline}).  
Once a \textsc{Shape} model, with a suitable morphology, density and velocity have been created it is then placed at an inclination of 90\degree. This 3D model must be represented two dimensionally,  by taking a 2D slice of one quarter of the model.  A data cube is created that describes the velocity and density at each position in the shell, this is processed by \textsc{PyCross}.
A series of 1D \textsc{Cloudy} simulations are computed along the 2D \textsc{Shape} model slice. Lastly, the 2D photoionisation map is wrapped and flipped in order to create the full pseudo 3D photoionisation model.   Currently this technique is constrained to axisymmetric nebulae, but as discussed later this allows for the modelling of a large majority of PNe and novae.

The main GUI allows users to set the name of the current model, blackbody temperature, total luminosity, angle range (start-finish) and the number of angles in the range.  \textsc{Cloudy} preferences for the model general input, i.e. options, abundances and emissions are entered into smaller windows and when saved, will remain when the programme is started again.  There is no limit to the number of parameters entered into these windows provided that they conform to valid \textsc{Cloudy} commands.  This removes any learning curve, complex commands and the need to run shell scripts, thus increasing productivity.  Once a model is successfully created the emissions list, set in the main GUI will be available and visible for selection in the plotting option GUI.  Button functionality on the main GUI is as follows:

\begin{itemize}
	\item \textit{\textbf{Select Shape File:}} Allows the user to select the .txt data cube output from \altshape{} (see \autoref{shapetextfile}).
	\item \textit{\textbf{Cloudy Options:}} Opens up a new window allowing the user to type/paste \altcloudy{} options, visible at the top left of \autoref{fig:GUI_Overview}. For example when creating a model for a nova, options might include:\\ \\
	\textit{``cosmic ray background\\
	element limit off -8\\
	age 8114 days"}\\
	\item \textit{\textbf{Abundances:}} Opens a new window allowing the user to type/paste abundances to be used as input to the model.  Examples of abundances can be found by clicking \textit{\textbf{Reference \(|\)  Abundance Examples}} in the menu bar.
		\item \textit{\textbf{Emissions:}} Opens a new window allowing the user to type/paste emissions to be used as input to the model.  Emissions must be added in a particular format of no more than 10 characters. To ensure that emissions are entered correctly, \altpycross{} has an extensive library of emission lines that is available by clicking \textit{\textbf{Reference \(|\)  Emission Line Index}} on the menu bar.
		\item \textit{\textbf{Open Archive Folder:}} Opens a new Finder/Explorer window where all \altpycross{} data is stored.
		\item \textit{\textbf{Run \altpycross{} - New \altshape{} Model:}} This button is clicked to run a new model; the user must first have selected a  \altshape{} data cube and entered all necessary \altcloudy{} options.   
		\item \textit{\textbf{Run \altpycross{} - Archive \altshape{} Model:}} If users create a model and are not satisfied with the outcome or feel that certain parameters need to be changed then they can modify specific \altcloudy{} options outlined above. By clicking this button and selecting an archived model, users can run the model again with updated parameters. This feature saves a lot of time as it does not require the user to start anew.
		\item \textit{\textbf{Plot Current Model:}} Open the plotting GUI to plot the current \altpycross{} model.  Any emissions entered will automatically be loaded into the plotting options window.
		\item \textit{\textbf{Plot Archive Model:}} Select a zipped archived model then open the plotting GUI and proceed to plot.  This again saves time not having to create a model from the start, while also offering the user the choice of building a database of models to plot from.
\end{itemize}

An $\lq$emission' as discussed here is a simulated narrow-band image at the wavelength of a specific spectral emission line. Users can plot from one to a maximum of six emissions (plotted in flux units ergs $\mathrm{s^{-1}}$) in a single plot by highlighting desired emission, and adding them to the ``\textit{Selected Emissions}" list.  Once selected, a plotting grid can be adapted to fit the required number of emissions.  The number of angles, inclination and labels for x-axis and y-axis are set prior to plotting.  Corresponding emission lines will also be displayed at the top of each plot regardless of the number of plots/subplots, the effective temperature of the ionising source is located in the colour-bar to the right of each model.

Users can view a plot $\lq$cut' with a weighted by N\textsc{ii} of  N$^+$/O$^+$/N/O, which is automatically generated.  These weighted plots are adapted from the tutorial for \altpycloudy{} \citep{2013pyCloudy}\footnote{\url{https://sites.google.com/site/pycloudy/examples/example-3?authuser=0}}, where one of the foundations that \altpycross{} was developed on and show the ionised fraction versus neutral fraction for the two most commonly studied astrophysical metals, i.e. O and N.  Selection of the cuts is performed by first selecting an element, then its ionising state; a maximum of three cuts is allowed per plot. Users may open the plotting folder directly from this GUI and automatically archive all plots generated in that session. This is achieved by selecting the \textbf{\textit{Auto Archive}} checkbox; archive plots are saved using the name and time-stamp of creation similar to saving of newly generated models discussed earlier. 

The latest version of \altpycross{} includes the functionality that gives users the ability to plot in 71 different colour maps and magnify the centre regions of nebulae, allowing to better visualise regions and/or possible hidden features, see  \autoref{fig:colourMapOptions} and \autoref{fig:zoom}.  On analysis of \autoref{fig:colourMapOptions}, the outer regions of the lobes are more clearly defined in (a) than (b) or (c) where the following colour maps were applied:   (a) nipy\_spectral\_r, (b) gist\_stern\_r and (c) gist\_heat\_r\footnote{The r signifies that this particular colour map has been revised in matplotlib \url{https://matplotlib.org/3.1.1/gallery/color/colormap_reference.html}}.  Colour maps may be misleading, for example differences of regions within the outer lobes of (a) and (b) in \autoref{fig:colourMapOptions} are greatly exaggerated as compared to that in (c) which is more realistic.  Offering this feature allows users to find a trade off for investigating features of nebulae.  

The axes on all plots are normalised distance units (0-100), the colour-bar corresponds to the effective temperature of the ionising source, flux units in ergs s-1.  Users can add additional text to the axis labels via the Plotting Options GUI, see \autoref{fig:GUI_Overview}.  Mathematical notation can also be written for the plot axis labels when using a subset TeX\footnote{\url{https://matplotlib.org/3.1.1/tutorials/text/mathtext.html}} markup by placing text inside a pair of dollar signs (\$).

\begin{figure}[]
	\centering
	{\includegraphics[width=\columnwidth]{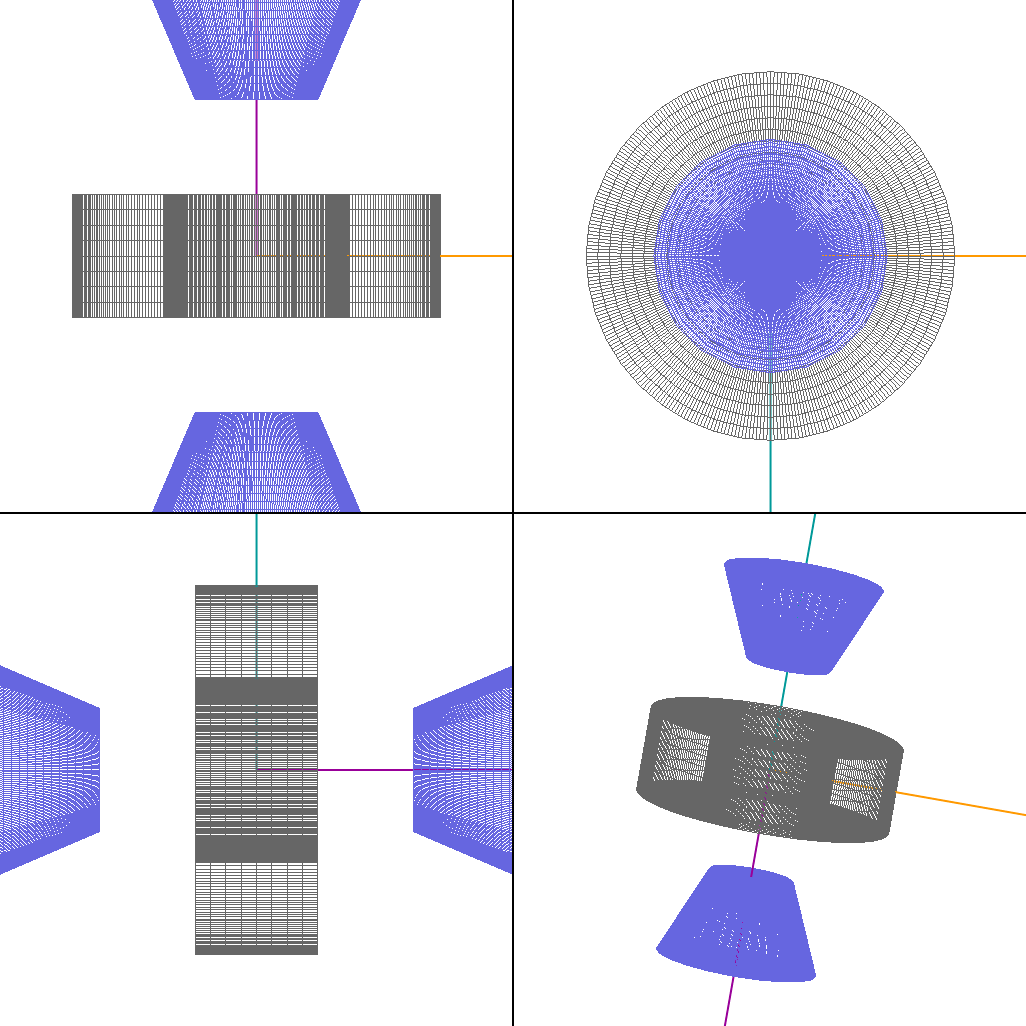}} %
	\caption[2018 Shape model for  V5668 Sgr]{\altshape{} model for  V5668 Sgr.   Figure adapted from \cite{2018Harvey_etal}}
	\label{fig:2018V5568}
\end{figure}

\begin{figure*}[t]
	\centering
	{\includegraphics[width=\textwidth]{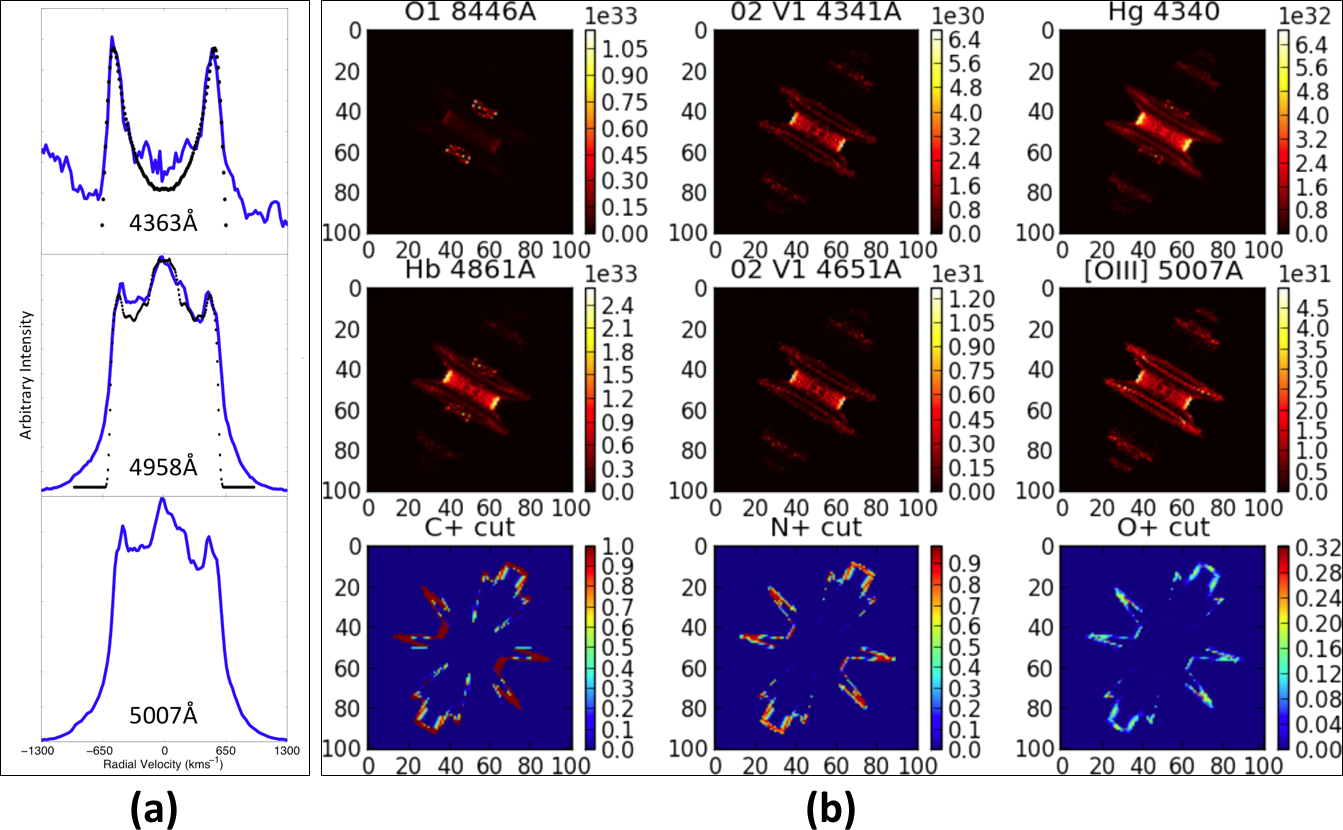}} %
		\caption[Model for Nova V5668 Sgr]{(a) Spectral line profile simulations of [O\,\textsc{iii}] nebular and auroral lines of V5668 Sgr on day 822 post discovery. The blue-solid lines represent observed line profiles and were used in the fitting of a morpho-kinematical model with the \textsc{\altshape{}} software, represented by overlaid black dots. The auroral line is fitted with an equatorial disk whereas the nebular lines fit an equatorial waist and polar cones morphology with a Hubble outflow velocity law. (b) {\altpycross} emission models of V5668 Sgr, using nova abundances with the inclination angle set at 85\degree. Clumpiness was simulated using a Perlin noise modifier in \textsc{\altshape{}}. The geometry used for this model is illustrated in \autoref{fig:2018V5568}. Figure adapted from \cite{2018Harvey_etal}, OII 4341$\AA$  was shown to demonstrate that the line emission is expected to arise from the same geometrical location and so to deblend it with 4340$\AA$ the same line profile shape would be assumed. With regard to the two nebular [O\,\textsc{iii}] lines they were shown in \cite{2018Harvey_etal} to show the consistency in the line shape as would be expected from 4959$\AA$ and 5007$\AA$ [O\,\textsc{iii}] and how different the auroral [O\,\textsc{iii}] line shape was.  Colour-bars correspond to the effective temperature of the ionising source, flux units in ergs $\mathrm{s^{-1}}$.  {\altpycross} scientific pipeline is illustrated in \autoref{fig:pipeline}
	\label{fig:ShapeV5668}}
\end{figure*}

\section{\textbf{\textsc{PyCross}} pipeline applied to novae}\label{Section3Novae}
Novae are the result of an eruption on the surface of a white dwarf in a close binary system. The white dwarf's counterpart is normally a red dwarf or sub-giant, which has overfilled its Roche lobe and thus loses mass through an accretion stream onto the white dwarf surface. Pressure at the white dwarf-accreted envelope interface increases due to a buildup of hydrogen-rich material, eventually resulting in thermonuclear runaway.  This subsequent eruption reaches luminosities $\gtrsim 10^4~\mathrm{{\rm L}_\odot}$, and ejects a mixture of material. A combination of that processed in the thermonuclear runaway, material dredged up from the white dwarf and the previously accreted outer layers of its companion. The ejected material reaches velocities of order $10^3$--$10^5\,\rm{km\,s^{-1}}$.  Emission line profiles indicate considerable spatial density and velocity structure.  Immediately after a nova eruption the ejected material is dense, bright and optically thick. This soon fades after revealing  H\,\textsc{i}  and He\,\textsc{i} emission lines.  Over time [O\,\textsc{iii}], [N\,\textsc{ii}] and [Ne\,\textsc{iii}] become stronger relative to the fading continuum \citep{2012AJ....144...98W}.  Over a few years the ejecta will be observed as a fading, constant-velocity expanding nebulous shell surrounding the post-nova star.  These eruptions are not destructive enough to change either star and generally they return to their quiescent state, on decadal timescales. Classical novae repeat the process every $\sim\,10^4\,\mbox{--}10^5\,\rm{yr}$ \citep{2006AstroBook}. Although, a recurrent nova population also exists, with observed recurrence periods on human timescales. Shorter rates of recurrence are related to heightened accretion rates onto higher mass white dwarfs \citep{1995ApJ...445..789P,2005ApJ...623..398Y,2010ApJ...725..831S}.
The material ejected from the white dwarf surface generally forms an axisymmetric shell of gas and dust around the system. These 3D shell structures are difficult to untangle as viewed on the plane of the sky without additional velocity information.  

While spectroscopic data can be used to yield approximate values for temperature, velocity, and density along the line of sight of the object, a photoionisation model is required to determine the chemical structure of a nebula \citep{bohigas2008photoionization}.  As discussed in \cite{2018Harvey_etal}, photoionisation modelling of ejected nova shells during their nebular stage of evolution can contribute to estimates of the total mass and abundances of heavy elements ejected during nova events \citep{2011Atypical}. Examples of the ability to realise photoionisation models of novae, more specifically V5668 Sgr and V4362 Sgr, created by the \altpycross{} pipelines, are outlined in \autoref{V5668_Paper} and \autoref{PTB_42}.

\subsection{Nova: V5668 Sagittarii (2015)}\label{V5668_Paper}
\cite{2018Harvey_etal} investigated V5668 Sgr (2015), a slow-evolving extremely bright nova on the surface of a CO white dwarf. The nova event produced dust \citep{2018DustV5668}, and was classified to be of the DQ Her-type\footnote{Archetype for rich dust-forming slow novae, and historically significant following a major observed eruption in 1934, one of the first novae to be analysed with high-cadence spectroscopy observations where results were later used to classify nova spectra into 10/11 subclasses by \cite{1942McLaughlin}}. The V5668 Sgr nova event holds the record for longest sustained gamma ray emission from such an event \citep{2018V5668}.

A goal of this research is to better understand the early evolution of classical nova shells in the context of the relationship between polarisation, photometry, and spectroscopy in the optical regime.  Observations over five nights with the Dipol-2 instrument mounted on the William Herschel telescope (WHT) and the Royal Swedish Academy of Sciences (KVA) stellar telescope La Palma, yielded polarimetric data directly following the nova's dust formation episode.  While this nova shell is not yet resolvable with medium sized ground-based telescopes, \cite{2018Harvey_etal} utilised \textsc{PyCross} to model and  visualise the ionisation structure of V5568 Sgr.  The polarimetric and spectroscopic data revealed conditions present in the expanding nova shell allowing for the creation of a \altshape{} model. The proposed geometry consisted of an equatorial waist and polar cones, see \autoref{fig:2018V5568}.

Initial models consisted of broad parameter sweeps that were refined using \altpycloudy{} \citep{2013pyCloudy}.  This allowed examination of emission line ratios for the hot--dense--thick nova shell under study.  A cylindrical primitive was used to construct the equatorial waist, where a density, thickness and Hubble velocity law were applied.  The polar features were constructed using cone primitives. In this instance densities applied were estimations from \altcloudy{} simulations and velocity components were derived from measuring Doppler broadened characteristics of observed emission lines.  Emission lines in fast outflows of unresolved structures are governed by their velocity field and orientation towards the observer. Analysis techniques available in \altshape{}  allow users to disentangle projection effects. The line shapes modelled in \autoref{fig:ShapeV5668} (a) are displayed at an inclination of 85\degree, a polar velocity of $940\,\rm{km\,s^{-1}}$ and an equatorial velocity of $640\,\rm{km\,s^{-1}}$ at their maximum extensions. 

Emission models of V5668 Sgr, were created using optimum values derived from \altcloudy{}/\altpycloudy{} parameter sweeps.  The average density was found to be $\sim 1.0\times10^9\,\rm{cm^{-3}}$, luminosity and effective temperature were set to log($\rm{{L_ \odot }}$) = 4.36 and $1.8 \times {10^5}\rm{K}$ respectively.  To recreate the nova conditions on day 141 post discovery an inner and an outer radius were set to $3.2\times10^{14}\,\rm{cm}$ and $6.4\times10^{14}\,\rm{cm}$, respectively.   \autoref{fig:ShapeV5668} (b) illustrates the \altpycross{}\ emission models for V5568 Sgr, where a comparison of the locality of emission through the shell of the same species is presented in each column of three panels. These models show the ionic cuts for C, N and O, respectively.  An ionic cut is a slice of the ionised structure for a specific ionisation state of a species. \autoref{fig:ShapeV5668} presents the first state of ionisation for C, N and O. The colour-bar shows the ionised fraction for different geometrical locations of the appropriate ionisation state.

\cite{2018Harvey_etal} revealed variability in polarisation suggesting internal shocks in the nova outflow, supported by the presence of gamma-ray emission \citep{2018V5668}.  The position angle of this nova was determined using the polarimetry observations. Spectroscopy allowed for derivation of the physical conditions, including outflow velocity and structure, nebular density, temperature and ionisation conditions.  Photoionisation models generated from \altpycross{} gave further insight into the nova system as a whole. \cite{2018Harvey_etal} concluded that slow novae are regularly referred to ``nitrogen flaring", however based on their findings suggest that they are in fact more likely ``oxygen flaring".

\begin{figure*}[!t]
	\centering
	{\includegraphics[width=6.6in]{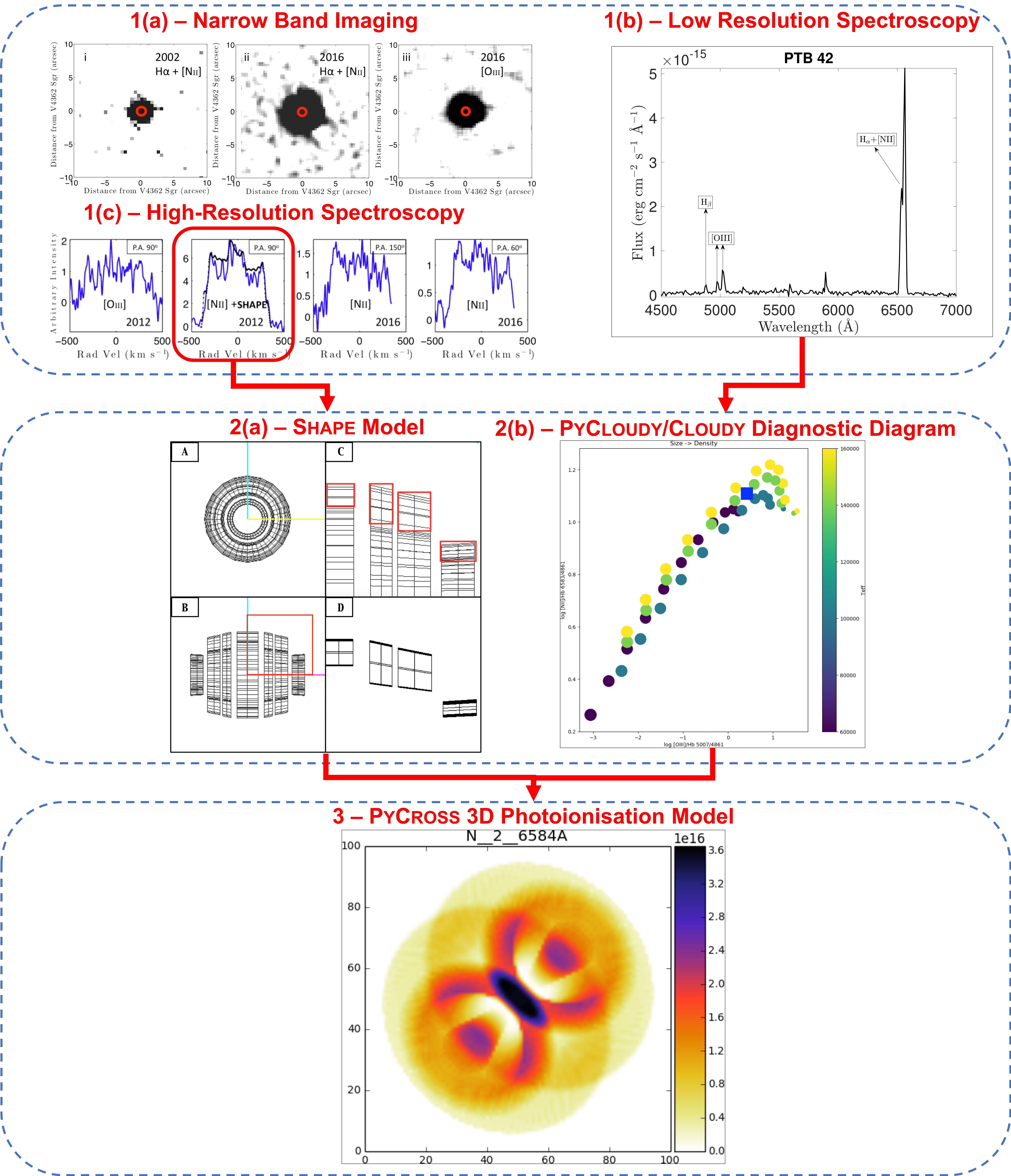}} %
		\caption[PyCross Scientific Pipeline]{\textsc{PyCross} Scientific Pipeline for V4362 Sgr\,/\,PTB 42(1994). This figure is split into three main sections (1-3). 1 represents observational data used to derive the spatial structure of an emission nebula, 1(a) is narrow-band imaging used to inform the extent and axial ratio of the nebula. 1(b) is low-resolution spectroscopy, used to derive density, abundances and ionisation conditions. 1(c) is high-resolution spectroscopy used to derive the velocity information for individual spectral lines. 2(a) is a suitable {\shape} geometry, seen in full in panels A and B, but arranged in panel D in a position to be output to a text file readable by \textsc{PyCross}, see section \ref{shapetextfile} and \cite{2020Harvey_etal}. 2(b) is a diagnostic diagram generated to interpret the emission line ratios measured from broadband low-resolution spectra. 3 is the final output \textsc{PyCross} model generated from applying the geometry and density at each geometrical point (i.e. from the \textsc{shape} output text file). Other parameters such as the stellar luminosity, stellar effective temperature, and abundances are input to {\pycross}'s GUI.  A set of 1D Cloudy simulations are run through the 2D parameter space which are then wrapped in azimuth around the complete shell creating a pseudo 3D model  for any observable emission line, here we show the [N\textsc{ii}] 6584$\AA$ at a position angle of 60\degree.  Colour-bars in 3 correspond to the effective temperature of the ionising source, flux units in ergs $\mathrm{s^{-1}}$, the colour-bar in the \pycloudy{} diagnostic diagram, 2(b), is the effective temp of the ionisation source. Figure adapted from \citep{2020Harvey_etal}.}
	\label{fig:pipeline}
\end{figure*}

\subsection{Nova: V4362 Sagittarii (PTB 42)}\label{PTB_42}
More recently \citep{2020Harvey_etal}, utilised the current version of {\pycross} to aid in uncovering a previously undiscovered classical nova shell surrounding the nova system PTB 42 of the DQ Her type.  Imaging was acquired from the Aristarchos telescope in Greece and consisted of two narrow-band filters;  H\,$\alpha\,+\,[N\textsc{ii}]$ (6578$\AA/40\AA$)\footnote{The first number represents the centre wavelength and second number represents the filter width in Angstroms} and [O\,\textsc{iii}] (5011$\AA/30\AA$) with exposures of 30\,-\,40 minutes in each filter.  High-resolution spectroscopic data was obtained using the Manchester Echelle Spectrograph (MES) instrument mounted on the 2.1\,m telescope at the San Pedro M\'artir (SPM) observatory, Mexico. The PTB 42 nova shell, was detected using the low-resolution, high-throughput SPRAT spectrograph on the Liverpool Telescope during August of 2016. The \altpycross{} pipeline (see \autoref{fig:pipeline}) was then used to generate emission models based on imaging and spectroscopic observations. Based on these observations,  the nova shell was reproduced in \altshape{} and then passed to \altpycross{}.

The P.A. for this target was chosen based on polarimetric observations presented in \cite{evanspol}, in addition to close examination of the major and minor axes in the narrow-band imaging (top left of \autoref{fig:pipeline}). High-resolution spectroscopy from MES was used to find the radial velocity of individual components within the nova shell.  Difficult to derive parameters are the system inclination, covering and filling factors of the nova shell, as well as the opening angle of the polar cones.

In this study a number of morpho-kinematic models were created to determine the best-fit relationship between image, PV array and 1D line spectrum to commonly proposed nova shells morphologies.  While spatial information could not be resolved, structures within the nova could be resolved via line-of-sight velocities. The observed equatorial velocity is 350\,km\,s$^{-1}$, as measured from the MES spectra. An axial ratio of 1.4 and polar velocity of 490\,km\,s$^{-1}$ was initially chosen based on the inclination corrected axial ratio for similar novae DQ Her and T Aur \citep{BodeNova}.  Adjusting for inclination when fitting to the line profile gives an equatorial velocity of 390\,km\,s$^{-1}$ and polar velocity of 550\,km\,s$^{-1}$.  This allowed for the remaining velocities to be set to 550$\,\times\,\rm{r/r0}$\, (km s$^{-1}$).  The observed [N~\textsc{ii}] line profile, taken from the 2012 observation (seen encircled in red at the top left of \autoref{fig:pipeline}, was chosen for modelling the nova shell structure as it had the highest S/N.

The luminosity of the system was estimated based on that of the class archetype, namely DQ Her, during quiescence \cite{1984Ferland}. The inner and outer radii of the shell are estimated based on the observed expansion velocity distribution and narrow-band imaging and shell age, although the actual shell thickness is difficult to know without resolving it spatially. Abundances used were for that of the archetypical slow nova, DQ Her \citep{1984Ferland}.

\begin{figure}[!ht]
\centering
\includegraphics[width=2.5in]{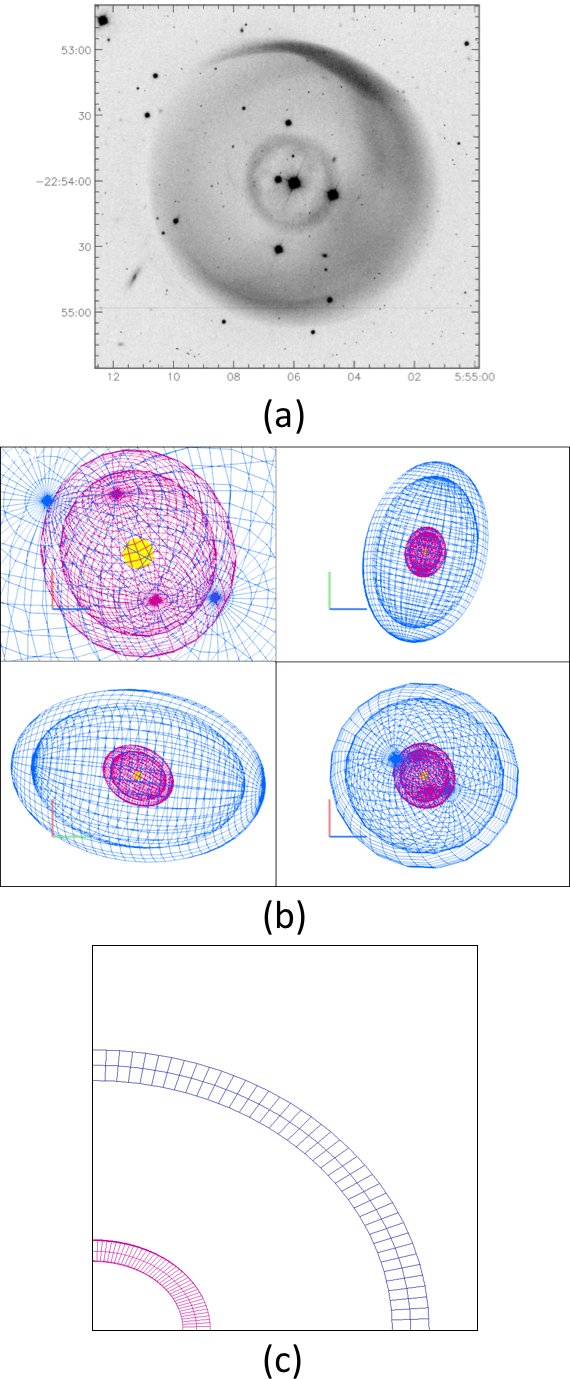}
\caption[LoTr 1 \textsc{SHAPE} Model Mesh (3D Module)]{(a) Deep, narrowband images of PN LoTr\,1, in [O\,\textsc{iii}] 5007$\AA$.   North is to the top of the image, East is to the left.  Image acquired using European Southern Observatory (ESO) Multi-Mode Instrument (EMMI),  mounted on the 3.6m New Technology Telescope (NTT) of La Silla Observatory. The outer shell appears brighter in both the northeast and southeast directions of the image, which could suggest the possibility of an inclined, elongated structure due to the projection effect.  \textit{ Image credit and \copyright \,} \cite{tyndall_lotr_2013}. (b) 3D \altshape{} model of LoTr1 from different views.  The view in the bottom right quadrant of (b) represents the nebula as in (a).  (c) \altshape{} quadrant slit of  LoTr 1 where shells have being aligned.  The resulting data cube is then passed to \altpycross{}.}
\label{fig:lotr1_Shape}
\end{figure}

In this work \altpycross{}  was employed to better understand observed ionised nebula structure by using broadband spectra from which line ratios are measured.  A grid of models was generated to determine the best fitting model parameters which were then applied through the derived geometry. This geometry is determined from matching line profiles in high-resolution spectra and narrow-band imaging in \altshape{}. Polarimetry and imaging was used to determine the position angle of the shell. Using abundances adapted from the DQ Her nova shell model of \cite{1984Ferland} \altpycross{}  generated a pseudo 3D simulation of the ionisation structure of PTB 42 as seen in top centre of \autoref{fig:pipeline}. Results show the difference in emission regions for the strongest nebular lines, i.e. [N \textsc{ii}] and [O \textsc{iii}].  The \altpycross{} model presented in \citep{2020Harvey_etal} detailed observed components i.e. equatorial ring, higher latitude rings and polar features of the nova, allowing their individual behaviour to be examined.

\section{\textbf{\textsc{PyCross}} pipeline applied to PNe}\label{Section4PNe}
Post-main sequence low to intermediate mass stars ($\sim0.8\,$M$_\odot-8\,$M$_\odot$) will over time evolve to become planetary nebulae as the outer layers of the star are ejected through thermal pulses after the AGB phase. This exposes hotter layers that can ionise the previous ejected material. A detailed description of the evolutionary track and mechanics a star undergoes on its journey to become a PN at each phase is presented in \cite{2009Prialnik} and see also  \citep{Herwig2005}. 

There are complications to this simple picture, particularly in the role of binarity and the possibility of `mimics' such as symbiotic systems \citep{boffin2019importance}. As a result, there is no universal observational definition for PNe, but their existence is usually dependent on two components: a circumstellar shell of sufficient mass and density to be detectable, and a hot central star to ionise it. PNe candidates are usually discovered by objective-prism surveys or by direct imaging in a narrow spectral region around a strong emission line or line such as [O\,\textsc{iii}] 4959$\AA$, 5007$\AA$  or H\,$\alpha$ and [N\,\textsc{ii}] 6548$\AA$, 6583$\AA$ \citep{2006AstroBook}.  Their expanding shell sizes ($\sim$\,0.2\,pc) and expansion velocities ($\sim$\,25\,km\,s$^{-1}$) imply a dynamical lifetime of $\sim$\,10$^4$\,yr \citep{kwok_2000}.

As intermediate mass stars make up a large portion of the stellar mass in our galaxy, studying how their nuclear-processed interiors are ejected into intricate nebulae and eventually into the interstellar medium, can lead to a deeper understanding of the galaxy's chemical evolution. PNe spectra are rich in emission lines, including the interesting \textit{forbidden lines}, and serve as a laboratory for the physics and chemistry of photoionisation. Modelling the 3D spatiokinematic structure, along with the 3D photoionisation of PNe and their mimics can contribute to understanding how these spectacular nebulae are formed.

\subsection{PN LoTr 1}
The previous section described \altpycross{} and its use in modelling and investigating novae emissions.  In this subsection, and for the first time, \altpycross{} will be used in modelling emissions for  LoTr\,1, a PN believed to contain a binary central star system consisting of a KL-type III giant and white dwarf, first discovered by \cite{longmore1980}.

A morpho-kinematic \altshape{} model (\autoref{fig:lotr1_Shape} (a)) was created based on long-slit Echelle spectroscopy acquired using the Anglo-Australian Telescope/University College London Echelle Spectrograph (AATUCLES) and the New Technology Telescope-ESO Multi-Mode Instrument (NTT-EMMI) focused on the [O\,\textsc{iii}] 5007$\AA$ emissions over eight different slit positions, published in \cite{tyndall_lotr_2013}.  Careful measuring of axes and relative sizes yielded a model constructed of two elongated shells with inner and outer shell radii of 5$''$ and 12$''$ respectively. Each shell is at different inclinations, with an angle of 50$\degree$ difference in position angle, and a difference of 57$\degree$ inclination (see \autoref{fig:lotr1_Shape} (a)). Modifiers were applied to elongate the inner and outer shells along the z-axis by a factor of 1.3 and 1.5 respectively. 

\begin{figure}[!ht]
\centering
\includegraphics[width=2.5in]{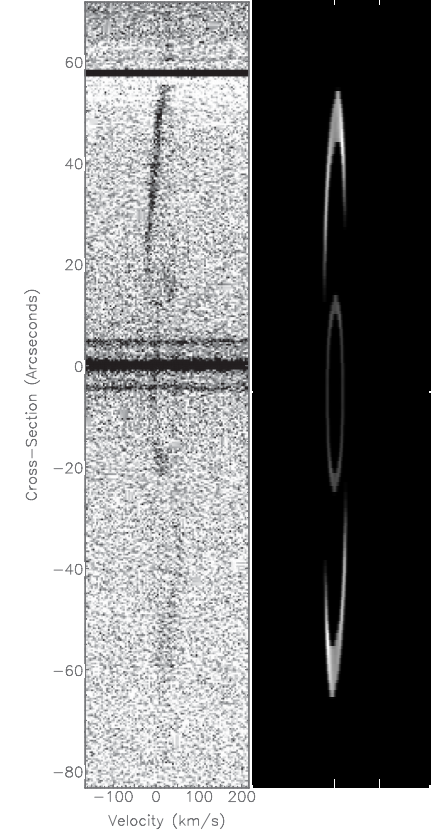}
\caption[LoTr 1 Slit 1 comparison]{Left: Recreation of slit position 1 acquired by \cite{tyndall_lotr_2013}.  Right: Theoretical PV diagram created in \altshape{}. West is up.}
\label{fig:slit1shape}
\end{figure}

\begin{table*}[!htbp]
	\centering
	\begin{tabular}{ccccc}
	\hline
	& \textbf{Radius (cm)}& \multicolumn{1}{c}{\textbf{\begin{tabular}[c]{@{}c@{}}Expansion Velocity \\ (km/s)\end{tabular}}} & \textbf{Morphology}  & \textbf{\begin{tabular}[c]{@{}c@{}}Kinematical Age \\ (yrs)\end{tabular}}\\ \hline
	\multicolumn{1}{c}{\textbf{Outer Shell}} & $2.84\times10^{16}$ & 25 $\pm$ 4& Elliptical& 35,000 $\pm$7,000 \\
\multicolumn{1}{c}{\textbf{Inner Shell}} & $9.14\times10^{15}$ & 17 $\pm$ 4& Elliptical& \multicolumn{1}{c}{17,000 $\pm$ 5,500} \\ \hline
& \multicolumn{1}{l}{} &  & \multicolumn{1}{l}{} &                                                   
	\end{tabular}
	\caption{Table of parameters for both inner and outer shells of LoTr 1 from \cite{tyndall_lotr_2013}.}
	\label{tab:lotr1}
\end{table*}

A comparison of the first slit acquired by \cite{tyndall_lotr_2013} compared to that recreated in \altshape{} can be seen in \autoref{fig:slit1shape}. The observed slit position is in the left hand panel and the right panel shows the theoretical PV array.  The systemic velocity, V$_{\mathrm{sys}}$ of the central shell was determined to be 14 $\pm$ 4 km/s \citep{tyndall_lotr_2013}, for the observations, which result from the velocity axis of this data being relative to the V$_{\mathrm{sys}}$.  An important similarity correspond to the major axis in that it has the same length as the diameter of the inner shell. This contributes to the assumption that it is a closed, isolated structure rather than a bipolar nebula. A bipolar nebular structure viewed from above would not give this regular shape. These ellipses seen along the velocity axis appear symmetric in both images, in agreement with a spherical shell or elongated ovoids.

\begin{figure}[!h]	
\centering
\includegraphics[width=2.2in]{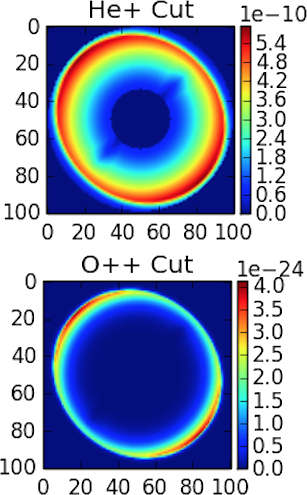}
\caption[LoTr1 \altpycross{} photoionisation Cuts]{LoTr 1 \altpycross{} photoionisation cuts of He$^{+}$ and O$^{++}$, using PN abundances with the inclination angle set at 40$\degree$.  Colour-bars correspond to the effective temperature of the ionising source, flux units in ergs $\mathrm{s^{-1}}$.}
\label{fig:pycross_cuts2}
\end{figure}

\begin{figure*}[!htb]  
\centering
\includegraphics[width=\textwidth]{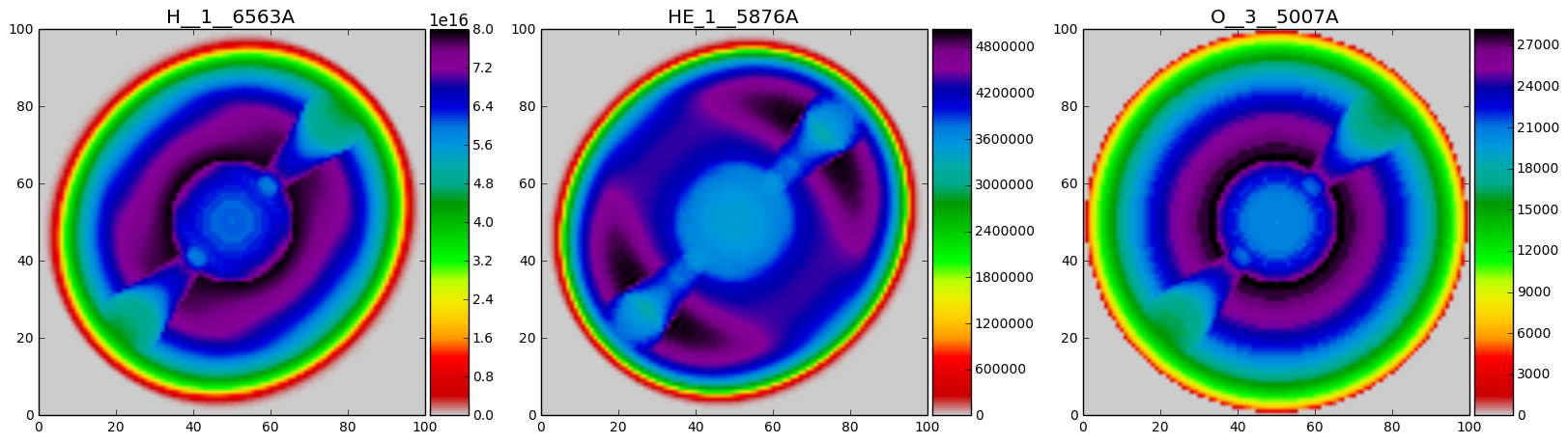}
\caption[LoTr1 \altpycross{} Photoionisation Model - Emissions 40$\degree$]{LoTr 1 \altpycross{} photoionisation model using PN abundances with the inclination angle set at 40$\degree$. The input \altshape{} model is seen in \autoref{fig:lotr1_Shape} with changes made to the inclination angles of the two shells to make the entire object symmetrical.  Colour-bars correspond to the effective temperature of the ionising source, flux units in ergs $\mathrm{s^{-1}}$.}
\label{fig:LoTr1_pycross_40}
\end{figure*}

Minor modifications were made to align the shells of LoTr1 prior to creating a data cube slice in order to create a photoionisation model using \altpycross{} (\autoref{fig:lotr1_Shape} (b)).  Default PN abudances were used from \cite{aller1983chemical} and \cite{khromov1989planetary}, with high depletions for elements not listed. Emission models generated from \altpycross{} for LoTr\,1 can be seen in \autoref{fig:LoTr1_pycross_40}.  These models are set at an inclination angle of 40$\degree$, with a blackbody effective temperature of $\sim$100,000 K \citep{gruendl2001variable} and a luminosity of 100 L$_{\odot}$ \citep{henry1999morphology}. These parameters were chosen as they were ascertained from the Helix nebula (NGC 7293), one of the closest and brightest PN \citep{hora2006infrared}. While a hot black body spectrum is used for the modelling here, it is possible to use more complex central star spectra such as, for example, WR-type spectra, as applicable to HB4 \citep{2019Sophia} and generally found in around 10\% of PNe \citep{2011DePew}. \autoref{fig:pycross_cuts2} shows ionic cuts of He$^{+}$ and O$^{++}$, top and bottom respectively, at an inclination angle of 90$\degree$ for the left column, and 40$\degree$ for the right column. It is worth noting that the object appears entirely spherical at 90$\degree$ because this angle shows a cut through the object head on. Similar observations were made by \cite{tyndall_lotr_2013} who reported clear asymmetry in the nebular structure, additionally but it is  close to a head on view.

In this scenario and for simplicity, attention is limited to two elements: oxygen and helium. While diffuse nebulae contain vast quantities of material, they have relatively low densities (approximately $10^{3}$ particles $\mathrm{cm^{-3}}$). They are rich with emission lines such as Balmer lines of hydrogen, as well as those which arise from the transitions between energy levels which result in ions such as O$^{+}$, O$^{++}$, N$^{+}$, etc.  As the lifetime of ionised hydrogen is very short (fractions of a microsecond due to the fast nature of electric dipole transitions), the probability of excited hydrogen is negligible, so any hydrogen can be assumed to be in its ground state for the purpose of ionisation rate calculations.

\autoref{fig:pycross_cuts2} compares O$^{++}$ and He$^{+}$ in our model for LoTr 1 showing predicted emissions maps of this PN in spectral lines other than those observed, to be compared with future observations. The stellar spectrum and nebular analysis could be developed further from this starting point based on more information regarding the central star.  This would require a more extensive data-set for further investigation into the morphology of LoTr 1 in order to confirm its true structure and inclination.  A data-set with a higher signal-to-noise ratio would be necessary to resolve the very faint outer shell.

Oxygen is a particularly useful ion in analysing the physical structure of a nebula. The well known [O\textsc{iii}] 5007$\AA$ optical line strongly contributes to nebula cooling, while more highly ionised OVI lines can radiate in at higher energies in hotter parts of the nebula. Where ionised hydrogen is found, we expect to find ionised oxygen as the first ionisation potential of oxygen is almost identical to that of hydrogen. The difference between the two however, is that oxygen can be ionised more than once \citep{dyson1997physics}.

\section{Conclusion}\label{Section5Conslusion}
A new application has been presented for the generation of photoionisation pseudo 3D modelling of thin shell nebulae modelled in \altshape{}.  Functionality, an operational overview and a scientific pipeline has been described with scenarios where \altpycross{} has been adopted for novae and PN.  The software was developed using a formal software development lifecycle,  written in Python and will work without the need to install any development environments or additional python packages.  This application, \altshape{} models and \altpycross{} archive examples are freely available to students, academics and research community on GitHub (\url{https://github.com/karolfitzgerald/PyCross\_OSX\_App}) for download.  The authors cordially request that this paper be referenced when using this tool. 

\section*{Acknowledgements}		
This work was supported by Athlone Institute of Technology and National University of Ireland-Galway and funded in part by the Irish Research Council's postgraduate funding scheme.
This development of this work was helped by invaluable discussions with Dr Christophe Morisset of UNAM. We also wish to acknowledge the databases used that made calculations in this work possible. Namely recombination coefficients were taken from \url{http://amdpp.phys.strath.ac.uk/tamoc/RR/} and \url{http://amdpp.phys.strath.ac.uk/tamoc/DR/} and the ionic emission data is from version 7.0 of CHIANTI. CHIANTI is a collaborative project involving the NRL (USA), the Universities of Florence (Italy) and Cambridge (UK), and George Mason University (USA).

\section*{References}

\bibliography{mybibfile}

\end{document}